\documentclass[proof]{WileyASNA-v1}
\usepackage{comment}

\def\bea{\begin{eqnarray}}
\def\beann{\begin{eqnarray*}}
\def\eea{\end{eqnarray}}
\def\eeann{\end{eqnarray*}}
\def\beq{\begin{equation}}
\def\ee{\end{equation}}
\def\eeq{\end{equation}}
\def\lan{\langle}
\def\ran{\rangle}
\def\3p0{$^{3}P_{0}$}
\def\aq{\bar{q}}
\def\nn{\nonumber}

\def\calh{{\cal H}}
\def\calh{{\cal H}}

\articletype{Article Type}%

\received{1 October 2020}
\revised{12 October 2020}
\accepted{25 October 2020}

\begin{document}

\title{Charmonium decay in the  C$^3$P$_0$ Model }

\author[1]{Ahmed Abd El-Hameed,  G. S. Hassan}

\author[2]{D. T.  da Silva}

\author[3]{J. N. de Quadros, D. Hadjimichef*}

\authormark{Ahmed Abd El-Hameed et al.}

\address[1]{\orgdiv{Physics Department }, 
\orgname{Assiut University}, \orgaddress{\state{ Assiut}, \country{Egypt}}}

\address[2]{\orgdiv{Instituto de F\'{\i}sica e Matem\'atica}, 
\orgname{Universidade Federal de  Pelotas (UFPel)}, \orgaddress{\state{ Pelotas, Rio Grande do Sul}, \country{Brazil}}}

\address[3]{\orgdiv{Instituto de F\'isica}, \orgname{Universidade Federal do Rio Grande do Sul (UFRGS)}, \orgaddress{\state{Porto Alegre}, \country{Brazil}}}

\corres{*Av. Bento Gon\c{c}alves, 9500 - Agronomia, Porto Alegre - RS, 91501-970.  \email{dimihadj@gmail.com}}

\abstract{
  In this work, we use the $C^3P_{\,0}$ model to
calculate the  decay widths of the
 low lying charmonium  $J^{PC}=1^{--}$ states, nominally  $J/\psi(1S)$ and $\psi(2S)$,
in the following common channels:
$\rho\,\pi$, $\omega\,\eta$, $\omega\,\eta^\prime$, $K^{\ast
+}\,K^-$, $K^{\ast 0}\,\bar{K}^0$, $\phi\,\eta$,
$\phi\,\eta^\prime$.  
 }

\keywords{ Charmonium decay; $C^3P_0$ model; Fock-Tani formalism.}

\maketitle

\section{Introduction}

 One of the main challenges of hadron physics today is the search for exotic excitations. In this direction the
 PANDA Experiment will be one of the future experiments to probe hadron matter with this goal. Located 
at the Facility for Antiproton and Ion Research (FAIR), PANDA is an acronym 
of antiProton ANnihilation at DArmstadt in Darmstadt, Germany \citep{panda}.

Many of the recent observations have been made in the charmonium energy regime. For example, the
$\chi_{c1}(3872)$, $\psi(4260)$ and $Z_c(4430)$  states  exhibit unexpected exotic properties. While the $\chi_{c1}(3872)$ and $Z_c(4430)$
could be multiquark states, the $\psi(4260)$ is a candidate for being a genuine $c\bar{c}g$ charmonium hybrid \citep{exotic-xyz}.
Concerning open charm and charm strange systems for many of the recent observations like e.g. the $D^\ast_{s0}(2317)^{\pm}$ or 
$D_{s1}(2460)^{\pm}$ the internal structure is still unknown.
 
The charmonium system, since its discovery in 1974  \citep{jpsi1,jpsi2}, 
 has become a reference in the study of meson spectroscopy. 
The experimentally clear spectrum of relatively narrow states below the open-charm
$DD$ threshold of 3.73 GeV can be identified with the $1S$,
$1P$, and $2S$ $c\bar{c}$  levels predicted by potential models, which
incorporate a color Coulomb term at short distances and a
linear scalar confining term at large distances \citep{barnes0}.

In the present work, we shall   concentrate  on the low lying charmonium   mesons,
which are the   $J/\psi(1S)$ and $\psi(2S)$ states.
We employ a mapping technique in order
to obtain an effective interaction for meson decay.
A particular mapping technique
long used in atomic physics, the Fock-Tani formalism
(FTf), which has been applied to hadron-hadron scattering interactions with constituent
interchange \citep{annals}. This technique has been extended in order to
include meson decay \citep{prd08}.
In summary,   starting with a microscopic $q\bar{q}$ pair-creation interaction,
in the leading order, the \3p0 results are reproduced. In the NLO  corrections appear due to
the bound-state nature of the mesons that   modifies
$q\bar{q}$ interaction strength, which is the C\3p0 model.

\section{ Fock-Tani Formalism and the  C$^3$P$_0$ model}
\label{sec:mesons}

This section reviews the formal aspects of the mapping procedure and
how it is implemented  to quark-antiquark meson states \citep{plb,annals}.
In the Fock-Tani formalism  one starts with the Fock representation of the
system using field operators of elementary constituents which satisfy canonical
(anti) commu\-ta\-tion relations. Composite-particle field operators  are
linear combinations of the elementary-particle operators and do not generally
satisfy canonical (anti) commutation relations. ``Ideal" field
operators acting on an enlarged Fock space are then introduced in close
correspondence with the composite ones.
Next, a given unitary transformation, which transforms the
single composite states into single ideal states, is introduced.
Application of the
unitary operator on the microscopic Hamiltonian, or on
other hermitian operators expressed in terms of the elementary constituent
field operators, gives equivalent operators
which contain the ideal field operators. The effective Hamiltonian
in the new representation
has a clear physical
interpretation in terms of the processes it describes. Since all field
operators in the new representation satisfy canonical (anti)commutation
relations, the standard methods of quantum field theory can then be readily
applied.

The starting point is the definition of single
composite bound states.  We write a single-meson state in terms of a meson
creation operator $M_{\alpha}^{\dagger}$ as
\bea
|\alpha \ran  = M_{\alpha}^{\dagger}|0 \ran ,
\label{1b}
\eea
where $|0 \ran$ is the vacuum state. The meson creation operator $M_{\alpha}^{\dagger}$ is written
in terms of constituent quark and antiquark creation operators
$q^{\dagger}$ and $\aq^{\dagger}$,
\bea
M^{\dagger}_{\alpha}= \Phi_{\alpha}^{\mu \nu}
q_{\mu}^{\dagger} {\aq}_{\nu}^{\dagger} ,
\label{Mop}
\eea
$\Phi_{\alpha}^{\mu \nu}$ is the meson wave function and $q_{\mu}|0
\ran=\aq_{\nu}|0\ran=0$. The index $\alpha$ identifies the meson
quantum numbers of space, spin and isospin. The indices $\mu$ and
$\nu$ denote the spatial, spin, flavor, and color quantum numbers of
the constituent quarks. A sum over repeated indices is implied. 
It is convenient to work
with orthonormalized amplitudes,
\bea
\lan\alpha|\beta\ran= \Phi_{\alpha}^{*\mu \nu}
\Phi_{\beta}^{\mu \nu}=\delta_{\alpha \beta}.
\label{norm}
\eea
The quark and antiquark operators satisfy canonical anticommutation relations,
$\{q_{\mu}, q^{\dagger}_{\nu}\}=\{\aq_{\mu},\aq^{\dagger}_{\nu}\}=\delta_{\mu \nu}, $
and $\{q_{\mu}, q_{\nu}\}=\{\aq_{\mu},\aq_{\nu}\}=
\{q_{\mu}, \aq_{\nu}\}= \{q_{\mu}, \aq_{\nu}^{\dagger}\}=0$\,.
%
%
Using these quark anticommutation relations, and the normalization condition
of Eq.~(\ref{norm}), it is easily shown that the meson operators satisfy the
following non-canonical commutation relations
\beq
[M_{\alpha}, M^{\dagger}_{\beta}]=\delta_{\alpha \beta} -
M_{\alpha \beta},\hspace{1.5cm}[M_{\alpha}, M_{\beta}]=0,
\label{M-com}
\eeq
where
\bea
M_{\alpha \beta}= \Phi_{\alpha}^{*{\mu \nu }}
\Phi_{\beta}^{\mu \sigma }\aq^{\dagger}_{\sigma}\aq_{\nu}
+ \Phi_{\alpha}^{*{\mu \nu }}
\Phi_{\beta}^{\rho \nu}q^{\dagger}_{\rho}q_{\mu}.
\label{delta}
\eea
A transformation is defined such that  a single-meson state
$|\alpha \ran$ is redescribed by an (``ideal") elementary-meson state by
\bea
|\alpha \ran\longrightarrow U^{-1}|\alpha\rangle =
m^{\dagger}_{\alpha}|0\rangle,
\label{single_mes}
\eea
where $m^{\dagger}_{\alpha}$ an ideal meson creation operator. The ideal
meson operators $m^{\dagger}_{\alpha}$ and $m_{\alpha}$ satisfy,
by definition, canonical commutation relations
\beq
[m_{\alpha}, m^{\dagger}_{\beta}]=\delta_{\alpha \beta} ,
\hspace{1.5cm}[m_{\alpha}, m_{\beta}]=0.
\label{mcom}
\eeq
The state $|0\rangle$ is the vacuum of both $q$ and $m$ degrees of freedom in the
new representation.
In addition, in the new representation the quark and antiquark operators
$q^{\dagger}$, $q$, $\aq^{\dagger}$ and $\aq$ are kinematically independent of
the $m^{\dagger}_{\alpha}$ and $m_{\alpha}$
\bea
[q_{\mu},m_{\alpha}]=[q_{\mu},m^{\dagger}_{\alpha}]=[\aq_{\mu},m_{\alpha}]=
[\aq_{\mu},m^{\dagger}_{\alpha}]=0 .
\label{indep_mes}
\eea
The
unitary operator $U$ of the transformation is 
\bea 
U(t)=\exp\left[t\, F\right] , 
\label{u} 
\eea 
where $F$ is the generator of the
transformation and $t$ a parameter which is set to $-\pi/2$ to
implement the mapping. The generator $F$ of the transformation is
\bea 
F= m^{\dag}_{\alpha}\,\tilde{M}_{\alpha}-
\tilde{M}^{\dag}_{\alpha} m_{\alpha} 
\label{f-generator} 
\eea 
where
\bea 
\tilde{M}_{\alpha}=\sum_{i=0}^{n}\tilde{M}^{(i)}_{\alpha},
\label{mes_gen} 
\eea 
with 
\bea
&&[\tilde{M}_{\alpha},\tilde{M}^{\dagger}_{\beta}] =
\delta_{\alpha\beta}
\hspace{.5cm} + \hspace{.5cm}{\cal O} (\Phi^{n+1}),\nn\\
&&[\tilde{M}_{\alpha}, \tilde{M}_{\beta}]= [
\tilde{M}^{\dagger}_{\alpha},  \tilde{M}^{\dagger}_{\beta}]=0.
\label{comO} 
\eea 
It is easy to see from (\ref{f-generator}) that
$F^{\dag}=-F$ which ensures that $U$ is unitary. The index $i$ in
(\ref{mes_gen}) represents the order of the expansion in powers of
the wave function $\Phi$. The $\tilde{M}_{\alpha}$ operator is
determined up to a specific order $n$ consistent with (\ref{comO}).

The next step is to obtain the transformed operators in the new
representation. The basic operators of the model are expressed in
terms of the quark operators. Therefore, in order to obtain the
operators  in the new representation, one writes
\bea
q(t)=U^{-1}\, q \,U,\hspace{1.0cm}{\aq}(t)=U^{-1}\, {\aq}\, U .
\eea

Once a microscopic interaction Hamiltonian $H_I$ is defined, at the quark
level, a new transformed Hamiltonian can be obtained. This effective
interaction,  the {\sl Fock-Tani Hamiltonian} ($\calh_{\rm FT}$),  is
obtained   by the application of the unitary operator
$U$ on the microscopic Hamiltonian $H_I$, {\it i.e.}, $\calh_{\rm FT}=U^{-1}\,H_I\,U$.
The transformed Hamiltonian   describes all possible
processes involving mesons and quarks.
The general structure of $\calh_{\rm FT}$  is of  the following form
\bea
\calh_{\rm FT}=\calh_{\rm q} + \calh_{\rm m} + \calh_{\rm m q} ,
\label{separation}
\eea
where the first term involves only quark operators, the second one involves
only ideal meson operators, and $\calh_{\rm m q}$ involves quark and
meson operators.
In $\calh_{\rm FT}$ there are higher order terms that provide bound-state corrections
to the lower order ones. The basic quantity for these corrections is the {\it bound-state kernel} (BSK)
$\Delta$ defined as
\begin{eqnarray}
\Delta(\rho\tau;\lambda\nu)
=\Phi^{\rho\tau}_{\xi}\Phi^{\ast\lambda\nu}_{\xi}.
\label{kernel}
\end{eqnarray}
The physical meaning of the $\Delta$ kernel becomes evident, in the
sense that it   modifies  the quark-antiquark interaction strength \citep{annals,prd08}.

\begin{figure}[t]
\centering
\includegraphics[width=90mm,height=80mm]{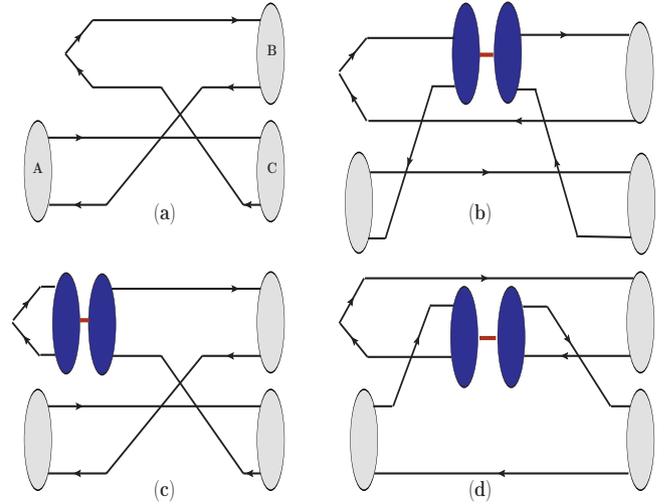} 
\caption{ Diagrams representing the decay  amplitude of the C\3p0 model, $h_{fi}^{\rm C3P0}$.
 The diagram (a) is the usual \3p0 ; diagrams  (b), (c)
and (d) correspond, respectively, to the three bound-state correction terms in  
Eq. (\ref{deltas}), where the dark blue bubbles represent the bound-state kernel $\Delta(\rho\tau;\lambda\nu)$ insertion.  } 
\label{c3p0-graf}
\end{figure}

In the present calculation, the microscopic interaction Hamiltonian
is a  pair creation Hamiltonian $H_{q\aq}$  defined as
\bea H_{q\aq}=V_{\mu\nu}\,
q^{\dag}_{\mu}\aq^{\dag}_{\nu} \,,
\label{h_3p0}
\eea
where in (\ref{h_3p0})
 a sum/integration is again  implied  over repeated indexes \citep{prd08}.
The pair creation potential $V_{\mu\nu}$ is given by
\bea
V_{\mu\nu}\equiv
 g\,
\delta_{c_{\mu}c_{\nu}}\delta_{f_{\mu}f_{\nu}}
\delta(\vec{p}_{\mu}+\vec{p}_{\nu})\,
\bar{u}_{s_{\mu}} (\vec{p}_{\mu}) \, 
v_{s_{\nu} }(\vec{p}_{\nu}) ,
\label{vmn}
\eea
with $g=2\,m_{q}\, \gamma\,$, where $\gamma$ is the pair production strength
and the indexes $c_\mu$, $f_\mu$, $s_\mu$ are of color, flavor and spin. 
The pair production is obtained from the  non-relativistic limit
 of $H_{q\aq}$ involving Dirac quark fields \citep{barnes1}.
Applying the Fock-Tani transformation to $H_{q\aq}$ one obtains the effective
Hamiltonian that describes a decay process.
In the FTf perspective a new
 aspect is introduced  to meson decay: bound-state corrections.
The lowest order correction is one that involves only one
bound-state kernel $\Delta$.
The bound-state corrected, C\3p0 Hamiltonian, is
\bea
\!\!\!\!\!\!\!\!\!\!\!\!\!\!\!\!
H^{\rm C3P0}
&=& -\Phi^{\ast\rho\xi}_{\alpha} \Phi^{\ast\lambda\tau}_{\beta}
\Phi^{\omega\sigma}_{\gamma}\, V^{\rm C3P0}\, m^{\dag}_{\alpha}
m^{\dag}_{\beta} m_{\gamma},
\label{c3p0}
\eea
where $V^{\rm C3P0} $
is a condensed notation for 
\bea
V^{\rm C3P0}&=&
\left(\,\bar{\delta}
+\bar{\Delta}\,\right)\,V_{\mu\nu}\,,
\label{vc3p0}
\eea
where
\bea
\bar{\delta}&=&\delta_{\mu\lambda}
\delta_{\nu\xi}
\delta_{\omega\rho}
\delta_{\sigma\tau}
\nn\\
\bar{\Delta}&=&
\frac{1}{4}\left[
\delta_{\sigma\xi}\,\delta_{\lambda\mu}\,\,
\Delta(\rho\tau;\omega\nu)
\frac{}{}
+
\delta_{\xi\nu}\,\delta_{\lambda\omega}\,\,
\Delta(\rho\tau;\mu\sigma)
\right.
\nn\\
&&
\left. \frac{}{}
-2\delta_{\sigma\xi}\,\delta_{\lambda\omega}\,\,
\Delta(\rho\tau;\mu\nu)
\right]\,.
\label{deltas}
\eea
The first term of (\ref{vc3p0}), involving 
$\bar{\delta}$ is the usual  \3p0 decay potential. The following $\bar{\Delta}$ term,   containing three  $\Delta$'s, is the 
bound-state correction to the potential.
In the ideal meson space the initial and final states involve only ideal
meson operators $|i\rangle=m^{\dag}_{\gamma}|0\rangle$ and
$|f\rangle=m^{\dag}_{\alpha}m^{\dag}_{\beta} |0\rangle $.
The C\3p0 amplitude is obtained by the following matrix element,
\bea
\hspace{-1cm}
\langle f | H^{\rm C3P0} | i \rangle &=&
\delta(\vec{P}_\gamma-\vec{P}_\alpha-\vec{P}_\beta)\, h_{fi}^{\rm C3P0}
\label{ideal-matrix}
\eea
shown in Fig. \ref{c3p0-graf}\!.
The $h_{fi}^{\rm C3P0}$ decay amplitude is combined with
relativistic phase space, resulting in the   decay width
\bea
\Gamma_{\gamma\to \alpha\beta} =2\pi\,P\,
\frac{E_\alpha\,E_\beta}{M_\gamma}
\int\,d\Omega\,|h_{fi}^{\rm C3P0}|^{\, 2}
\label{dif-gamma}
\eea
which, after integration in the solid angle $\Omega$, a  usual choice for the meson
momenta is made: $\vec{P}_\gamma=0$
($P=|\vec{P}_\alpha|=|\vec{P}_\beta|$).

\section{Results and Conclusions}
 
As a testing ground for the C\3p0 model in the charm sector, we shall 
calculate the decay widths of the lowest charmonium $1^{--}$ states, nominally  $J/\psi(1S)$ and $\psi(2S)$ 
 in the following  common channels:
$\rho\,\pi$, $\omega\,\eta$, $\omega\,\eta^\prime$, $K^{\ast
+}\,K^-$, $K^{\ast 0}\,\bar{K}^0$, $\phi\,\eta$,
$\phi\,\eta^\prime$.

It is well known that the $c\bar{c}$ content for these decay channels are  strongly  suppressed by the OZI  rule  
\citep{okubo,zweig2,zweig3,iizuka}. For the C\3p0 model to be applicable to  these  channels we shall introduce
 a flavor mixing scheme. The  $J/\psi(1S)$, $\phi$ and $\omega$ mesons  shall be considered as
the following mixture \citep{feldmann1}  
\bea
|\omega\rangle&=&|n\bar{n}\rangle-0.06|s\bar{s}\rangle-1.5\times10^{-3}|c\bar{c}\rangle\nn\\
|\phi\rangle&=&0.06|n\bar{n}\rangle+|s\bar{s}\rangle-0.9\times10^{-3}|c\bar{c}\rangle\nn\\
|J/\psi\rangle&=&1.5\times10^{-3}|n\bar{n}\rangle+0.9\times10^{-3}|s\bar{s}\rangle+|c\bar{c}\rangle\,.
\label{mix} 
\eea
A similar mixture occurs for pseudoscalar mesons $\eta$, $\eta^\prime$ and
$\eta_c$ \citep{feldmann2}
\bea
|\eta\rangle&=&0.77|n\bar{n}\rangle-0.63|s\bar{s}\rangle-0.006|c\bar{c}\rangle\nn\\
|\eta\,^\prime\rangle&=&0.63|n\bar{n}\rangle+0.77|s\bar{s}\rangle-0.016|c\bar{c}\rangle\nn\\
|\eta_c\rangle&=&0.015|n\bar{n}\rangle+0.008|s\bar{s}\rangle+|c\bar{c}\rangle\,.
\label{mix-eta}
 \eea
We shall assume that the $\psi(2S)$ has the mixing as $J/\psi(1S)$.

The spatial wave functions used are of type harmonic oscillator.
Therefore, we have as free parameters, beyond the coupling constant
$\gamma$,  the widths $\beta$'s of the Gaussians. The mesons used in
the bound-state kernel  were the $\phi$ and $\omega$ and are allowed
to mix as in Eq. (\ref{mix}). A particular notation is introduced
for the Gaussian widths for the BSK meson: $\beta_{\phi_\Delta}$,
$\beta_{\omega_\Delta}$, $\beta_{J/\psi_\Delta}$  and $\beta_{\psi_\Delta}$.
The model was adjusted in order  to minimize $R$, defined by
\bea 
R^2=\sum_{i=1}^{7} \left[a_{i}(\gamma,\beta)-1\right]^2 
\label{hiper} 
\eea 
with  $a_{i}(\gamma,\beta)= \Gamma_{i}^{\rm thy}(\gamma,\beta) /\Gamma_{i}^{\rm exp} $.

 After the simulation the best $J/\psi(1S)$ fit to the experimental data is achieved with
$\gamma=0.33$, $\beta_{J/\psi}= 0.182$ GeV, $\beta_\rho=0.55$ GeV,
$\beta_\pi= 0.55$ GeV, $\beta_\omega= 0.328$ GeV, $\beta_{\eta}= 0.348$
GeV, $\beta_{\eta^\prime}= 0.170$ GeV, $\beta_K= 0.4$ GeV,
$\beta_{K^\ast}=0.315$ GeV, $\beta_{\phi}=0.340$ GeV,
$\beta_{\phi_\Delta}=0.3$ GeV, $\beta_{\omega_\Delta}=0.6$ GeV and
$\beta_{J/\psi_\Delta}=0.3$ GeV. These results and the experimental
values are shown in table \ref{tab1}.

The best $\psi(2S)$ fit to the experimental data is achieved with
$\gamma=0.33$, $\beta_{\psi}= 0.042$ GeV, $\beta_\rho=0.55$ GeV,
$\beta_\pi= 0.55$ GeV, $\beta_\omega= 0.7$ GeV, $\beta_{\eta}= 0.7$
GeV, $\beta_{\eta^\prime}= 0.8$ GeV, $\beta_K= 0.47$ GeV,
$\beta_{K^\ast}=0.8$ GeV, $\beta_{\phi}=0.7$ GeV,
$\beta_{\phi_\Delta}=0.9$ GeV, $\beta_{\omega_\Delta}=0.6$ GeV and
$\beta_{\psi_\Delta}=0.042$ GeV. These results and the experimental
values are shown in table \ref{tab2}.

\begin{table}[t]
\centering
\begin{tabular}{ccccc}
\hline
  $\Gamma$ &C3P0   & Exp 
\\
      & (keV)& (keV)
\\
%
   \hline
 $\rho\,\pi$              & 1.55 & 1.57     \\
  $\omega\,\eta$          & 0.16 & 0.16   \\
  $\omega\,\eta^{\,\prime}$   &0.17 & 0.02 \\
  $K^{\ast+}\,K^-$         & 0.47 & 0.46\\
  $K^{\ast 0}\,\bar{K}^0$  & 0.48& 0.39 \\
  $\phi\,\eta$            & 0.08 & 0.07 \\
  $\phi\,\eta^\prime$      & 0.04 & 0.04 \\
   \hline
 $R = 0.41$\\
\end{tabular}
\caption{Decay width of $J/\psi(1S)$ ,where 
the experiemental data is from PDG \citep{pdg}.
}
\label{tab1}
\end{table}

\begin{table}[t]
\centering
\begin{tabular}{ccccc}
\hline
  $\Gamma$ &C3P0   & Exp 
\\
      & (eV)& (eV)
\\
   \hline
 $\rho\,\pi$              & 9.47 & 9.40     \\
  $\omega\,\eta$          & 3.67 & 3.23   \\
  $\omega\,\eta^{\,\prime}$   &2.23 & 9.40 \\
  $K^{\ast+}\,K^-$         & 5.06 & 5.12\\
  $K^{\ast 0}\,\bar{K}^0$  & 5.09& 32.04 \\
  $\phi\,\eta$            & 1.51 & 9.11 \\
  $\phi\,\eta^\prime$      & 3.28 & 4.52 \\
   \hline
 $R = 1.44$\\
\end{tabular}
\caption{Decay width of $\psi(2S)$, again the experiemental data is from PDG \citep{pdg}.
}
\label{tab2}
\end{table}

This work is intended as an initial and modest guide for future experimental
studies of charmed meson spectroscopy, that may  become possible with
the  advent of  PANDA Experiment,  that will probe   the $c\bar{c}$ sector
in search of exotic states such as the $c\bar{c}g$ charmonium hybrid. 
  
The approach presents   improvements  when compared to the  leading order \3p0 model. 
A detailed comparison and discussion  of the \3p0 model with C\3p0 
can be found in \citep{prc20}.
It was shown that for $\phi$ mesons, which are strange quarkonia $s\bar{s}$ states,
the inclusion of the   correction terms reduced the $R$ value as a clear demonstration that 
the bound-state correction globally improves the fit.
The average difference between the predictions of \3p0 and C\3p0, in each individual channel, ranged from 1\% to 14\%.
The higher differences were in the channels with lighter mesons in the final state. In the heavier channels, the leading
order \3p0 is dominant and the bound-state corrections represent an actual {\it next to leading order} correction.
A   similar effect is expected  in the higher $c\bar{c}$ sector
and for exotic charmonium hybrid $c\bar{c}g$ states.

\bibliography{hadjimichef_iwara2020}

\end{document}